\documentclass[a4paper,11pt,twoside]{preprint}

\usepackage{times}
\usepackage{amsmath}
\usepackage{amssymb}
\usepackage{enumerate}
\usepackage{fancyhdr}
\renewcommand\refname{References}
\newcommand{\Oun}{\mathcal{O}(1)} 
\newcommand{\WW}{\mathcal{W}} 
\newcommand{\DD}{\mathcal{D}} 

\usepackage{amsthm}
\usepackage{mathrsfs}
\let\rho=\varrho
\let\phi=\varphi
\def\real{\mathbb R}
\def\varg{{\textstyle{\frac{p-\rho q}{1+\rho }}}}

\newtheorem{theorem}{Theorem}[section]
\newtheorem{lemma}[theorem]{Lemma}
\newtheorem{proposition}[theorem]{Proposition}

\def\ie{{\it i.e.}}
\def\HALF{{\textstyle\frac{1}{2}}}
\newcommand{\proba}{{\bf P}}

\newenvironment{remark}{\noindent\textbf{Remark.}}{}
\def\const{{\rm const. }}
\def\supp{{\rm supp}}

\newcommand{\pp}{\mathscr{P}}

\newcommand{\vbanach}{\mathscr{F}}
\newcommand{\vcone}{\mathcal{C_\vbanach}}
\def\HALF{{\textstyle\frac{1}{2}}}
\def\FOUR{{\textstyle\frac{1}{4}}}

\newcommand{\flpp}[1]{f^{+}_{{\rm L}, #1}}
\newcommand{\flmm}[1]{f^{-}_{{\rm L}, #1}}
\newcommand{\flp}{f_{\rm L}^{+}}
\newcommand{\flm}{f_{\rm L}^{-}}
\newcommand{\frp}{f_{\rm R}^{+}}
\newcommand{\frm}{f_{\rm R}^{-}}

\newcommand{\frmm}[1]{f_{{\rm R},  #1}^{-}}
\newcommand{\frpp}[1]{f_{{\rm R},  #1}^{+}}

\newcommand{\pdm}{{\textstyle{\frac{p}{m}}}}
\newcommand{\apdm}{{\textstyle{\frac{|p|}{m}}}}
\newcommand{\atpdm}{{\textstyle{\frac{|\tilde p|}{m}}}}
\newcommand{\garg}{{\textstyle{\frac{P-(1+\rho )p}{\rho}}}}

\newcommand{\gargf}{{\textstyle{\frac{\tilde p + \rho p}{1-\rho}}}}

\newcommand{\mdp}{{\textstyle{\frac{m}{p}}}}
\newcommand{\amdp}{{\textstyle{\frac{m}{|p|}}}}
\newcommand{\flux}{a}
\newcommand{\ope}{\mathscr{L}}

\newcommand{\OO}{{\cal{O}}}
\newcommand{\domain}{\mathcal{D}}
\newcommand{\voisin}{\mathscr{V}}

\newcommand{\banach}{\mathscr{G}}

\def\ap{p^2}
\def\aq{q^2}

\def\ttt#1{{\textstyle{#1}}}

\def\emph#1{{\it #1}}

\def\d{{\rm d}}

\let\epsilon=\varepsilon

\def\d{{\rm d}}
\def\YY{{\mathscr{Y}}}

\newcommand{\sign}{\text{sign}}
\def\eref#1{(\ref{#1})}
\usepackage{MHequ}

\begin{document}
\title{A Model of Heat Conduction}
\bigskip

\author{P. Collet${}^1$ and J.-P. Eckmann${}^2$}
\institute{${}^1$Centre de Physique Th\'eorique, CNRS UMR 7644,\\ Ecole
  Polytechnique, F-91128 Palaiseau Cedex (France)\\
\\
${}^2$D\'epartement de Physique Th\'eorique et Section de Math\'ematiques,
Universit\'e de Gen\`eve, CH-1211 Gen\`eve 4 (Switzerland)}

\maketitle

\thispagestyle{empty}
\begin{abstract}
We define a deterministic ``scattering'' model for heat conduction
which is \emph{continuous} in space, and which has a Boltzmann type
flavor, obtained by a closure based on memory loss between
collisions. We prove that this model has, for stochastic driving
forces at the boundary, close to Maxwellians, a unique non-equilibrium
steady state. 

\end{abstract}

\section{Introduction} 

In this paper, we consider the problem of heat conduction in a
model which is a \emph{continuous} approximation of a discrete model
of a chain of cells, each of which contains a (very simple) scatterer
in its interior. Particles move between the cells, interacting with
the scatterers, but not among themselves, similar to the model put
forward in \cite{EY2}.
 
After a quite detailed description of this
scatterer-model, we will finally arrive in Eq.~\eref{eqstat} at a
formulation with a continuous space variable $x$ varying in
$[0,1]$. This approximation will be obtained by taking formally the
limit of $N\to\infty $ cells, but taking each cell of length $1/N$.
(The reader who is interested only in the formulation of the $x$-continuous
equation can directly skip to Eq.~\eref{eqstat}.) We then proceed to
show our main result, namely the existence of solutions to this
Boltzmann-like equation for initial conditions close to
equilibrium. Therefore, this will show that, although the model is
deterministic (except for the boundary conditions), with no internal
dissipation, it has a (unique) non-equilibrium steady state when driven
weakly out of equilibrium. The only approximations of the model are
the limit of $N\to\infty$, and a closure relation which models a loss
of memory between collisions.

\subsection{One cell}

To define the model,
we begin by describing the scattering process in one cell.
We begin with the description of one ``cell''. The cell is
1-dimensional, of length $2L$, and with particles entering on either
side. These particles have all mass $m$, velocity $v$ and momentum
$p=m v$. These particles do not interact among themselves.
Note that $v\in \real$ and more precisely, $v> 0$ if the
particle enters from the left, while $v<0$ if it enters on the right
side of the cell. In the center of the cell, we imagine a
``scatterer'' which is a point-like particle which can exchange energy
and momentum 
with the particles, but does not change its own position. (This
scatterer is to be thought of as a 1-dimensional variant of the rotating disks used in
\cite{EY2}.)
The
scatterer has mass $M$ and its ``velocity'' will be denoted by $V$.
The collision rules are those of an elastic collision, where
$\tilde v$ and $\tilde V$ denote quantities after the collision while
$v$, $V$ are 
those before the collision. In equations,
$$
\begin{array}{cc}
\tilde v&=-\rho v+(1+\rho)V~,\\
\tilde V&=(1-\rho) v+\rho V~,\\
\end{array}
$$
with
\begin{equ}\label{quot}
\rho=\frac{M-m}{M+m}~,\qquad \mu\equiv\frac{m}{M}=\frac{1-\rho }{1+\rho }~.
\end{equ}
Note that $\rho\in(-1,1)$, since we assume $m$ and $M$ to be finite
and non-zero.
If $\tilde v>0$, we say that the particle leaves the cell to the
right; if $\tilde v<0$, we say it leaves to the left.

For simplicity, we will assume $\rho
>0$, that is, $M>m$.
For the momenta, we get the analogous rules
$$
\begin{array}{cc}
\tilde p&=-\rho p+(1-\rho)P~,\\
\tilde P&=(1+\rho) p+\rho P~.\\
\end{array}
$$
Note that the matrix
\begin{equ}\label{Sv}
S\,=\,\left( 
\begin{matrix}
    -\rho &1-\rho\\
    1+\rho& \rho
\end{matrix}
  \right)
\end{equ}
has determinant equal to $-1$ and furthermore $S^2=1.$

We next formulate scattering in terms of probability densities (for
momenta) for just one cell.  We denote by $g(t,P)$ the probability
density that at time $t$ the scatterer has momentum $P\; (=M V)$ and we will
establish the equation for the time evolution of this function. To
begin with, we assume that particles enter only from the left of the
cell, with momentum distribution (in a neighboring cell or a bath)
$p\mapsto \flp(t,p)$, where $p=m v$. Thus, there are, on average,
$p\flp(t,p)\d p /m$ particles entering the cell (per unit of time) from
the left with momentum in $[p,p+\d p]$. Note that $\flp$ has support
on $p\ge0$ only, (indicated by the exponent ``$+$''); it is the
distribution of particles \emph{going to enter the cell} from the
left. Also note that the distribution of the momenta after collision,
\ie, before \emph{leaving} the cell, is in general not the same as
$\flp$.

Denote by $\pp(t)$ the stochastic process describing the momentum of the
scatterer. We have for any interval (measurable set) of momenta $A$,
for the probabilities $\proba$:
$$
\proba\big(\pp(t+\d t)\in A\big)=
\proba\big(\pp(t)\in A\;;\;\text{no\ collision\ in}\;[t,t+\d t]\big)
$$
$$
+\proba\big(\pp(t+\d t)\in A\;;\;\text{collisions\ occurred\ in}\;[t,t+\d t]\big)\;.
$$
We assume for simplicity that with probability one, 
only one collision can occur in an interval $[t,t+\d t]$. 
If there is a collision in $[t,t+\d t]$ with a particle of velocity $v=p/m>0$, this
particle must have left the boundary at time $t-\mdp L$ with momentum
$p$. Therefore,
$$
\proba\big(\pp(t)\in A\;;\;\text{a\  collision\ occurred\ in}\;[t,t+\d t]\big)\;
$$
$$
=\d t \int_{A}\d \tilde P\int_{\real^{+}} \kern-0.7em\d p\,
\delta\big(\tilde P-\rho P-(1+\rho)p\big)\,g(t,P)\,\pdm\flp(t-\mdp L,p)\,
\;.
$$
This equation neglects memory effects coming from the fact that a
particle may have hit the scatterer, bounce out of the cell and reenter
to hit again the scatterer. Similarly,
$$
\proba\big(\pp(t)\in A\;;\;\text{no\  collision\ occurred\ in}\;[t,t+\d t]\big)\;
$$
$$
=\left(1-\d t\int_{\real^{+}}\kern -0.8em \d p\, \pdm \flp (t-\mdp L,p)\,
\right) \int_{A}\d P\, g(t,P)\;.
$$

We immediately deduce the evolution equation,
\begin{equa}\label{01bg}
\partial_{t}g(t,P)=&- g(t,P) \int_{\real^{+}} \kern-0.8em\d p\,\pdm \flp (t-\mdp L,p)\,
\\
& +\frac{1}{\rho}
\int_{\real^{+}} \kern-0.8em\d p\,
g(t,\garg)\,\pdm\,\flp (t-\mdp L,p)\,\;.
\end{equa}
Note that this equation preserves the integral of $g$ over $P$, \ie,
it preserves probability.

This identity generalizes immediately to the inclusion of injection
from the right, with distribution $\frm$ having support in $p<0$. One gets
\begin{equa}\label{1bg}
\partial_{t}&g(t,P)=- g(t,P) \int_{\real} \d p\,  \apdm \left(\flp (t-\mdp L,p)+\frm (t+\mdp L,p)\right)\,
\\
& +\frac{1}{\rho}\int_{\real} \d p\,
g(t,\garg)\, \apdm\,\left(\flp (t-\mdp L,p)+\frm(t+\mdp L,p)\right)\;.
\end{equa}
In the stationary case, this leads to
\begin{equ}\label{flfr}
g(P) =
\frac{1}{\rho\lambda}
\int_{\real}\d p\, g(\garg)\,\apdm \,\left(\flp (p)+\frm(p)\right)~,
\end{equ}
where 
\begin{equ}\label{lambda}
\lambda
= \int_{\real} \d p\, \apdm \, \left(\flp (p) +\frm(p)\right)
\end{equ}
is the particle flux (see Sect.~\ref{s:flux} below).

It is important to note that the solution $g$ of Eq.~\eref{flfr} only
depends on the 
\emph{sum}: $f=\flp+\frm$, and thus, we can define a map
\begin{equ}
  f\mapsto  g_f~,
\end{equ}
where $g_f$ is the (unique) solution of Eq.~\eref{flfr}. It will be discussed
in detail in Sect.~\ref{s:gequ}.

We can also compute the distribution of the momenta of the particles
after collision.
We have
$$
\proba\big(\tilde p\in A\;;\;\text{a\  collision\ occurred\ in}\;[t,t+\d t])\;
$$
$$
=\d t \int_{A}\d \tilde p\int_{\real}\d P\,
\delta\big(\tilde p+\rho p-(1-\rho)P\big)\,g(t,P)\,\apdm f (t-\amdp L,p)\;.
$$
This particle reaches the left or right boundary of the cell
(according to the sign of $\tilde p$) after a time $mL/|\tilde p|$ (assuming the
scatterer is located in the center of the cell). Therefore, we
have for the ejection distributions $\flm$ (on the left) and $\frp$
(on the right):
$$
\atpdm\flm (t,\tilde p)=\frac{\theta(-\tilde p)}{1-\rho }\int_{\real} \d p\,
g(t-{\textstyle{\frac{m}{|\tilde p|}}}L,\gargf)\apdm\,f (t-{\textstyle{\frac{m}{|\tilde p|}}}L-\amdp L,p)\;,
$$
and
$$
\atpdm\frp (t,\tilde p)=
\frac{\theta(+\tilde p)}{1-\rho }\int_{\real}\d p\, 
g(t-{\textstyle{\frac{m}{|\tilde p|}}}L,\gargf)\apdm\,f
(t-{\textstyle{\frac{m}{|\tilde p|}}}L-\amdp L,p)\;,
$$
where $\theta$ is the Heaviside function.
In the stationary case we get
\begin{equ}\label{out1}
\atpdm\flm (\tilde p)=\frac{\theta(-\tilde p)}{1-\rho }\int_{\real} \d p\,
g(\gargf)\,\apdm f (p)~,
\end{equ}
and
\begin{equ}\label{out2}
\atpdm\frp (\tilde p)=\frac{\theta(+\tilde p)}{1-\rho }\int_{\real} \d p\,
g(\gargf)\,\apdm f (p)\;.  
\end{equ}
Since $g=g_f$ is determined by the incoming distribution $f_{\rm in}=\flp+\frm$
(and is unique if we normalize the integral of $g$ to 1)
\begin{equ}\label{intg1}
  \int_\real \d P\, g(t,P)=1~,
\end{equ}
we see that
the outgoing distribution $f_{\rm out}=\flm+\frp$ is entirely determined by the incoming distribution.
Note also that the flux is preserved:
$$
\int \d p\, \apdm f_{\rm in}(p)\,=\,\int \d p\, \apdm f_{\rm out}(p)~.
$$

\subsection{Stationary solutions for one cell}

Here, we will look for stationary states of the evolution equation
\eref{01bg}, which have also the property that the ejected
distributions are equal to the  injected ones. It is almost obvious
that Maxwellian fixed points can be found, but for completeness, we
write down the formulas. The reader should note that the distributions
$f_{\rm in}$ and $f_{\rm out}$ have singularities at $p=0$. This reflects the well-known fact
that slow particles need very more time to leave the cell than fast
ones. However 
$F(p)\equiv \pdm f(p)$ is a very nice function, and it is this function
which appears in all the calculations of the fluxes, and stationary
profiles. In this section we do the calculations with the quantity
$f$. Starting from Sect.~\ref{s:continuous}, we will use $F$.

We impose the two incoming distributions
$$
\flp (p)=\sigma \theta(+p)\amdp e^{-\beta p^{2}/(2m)}\;,
$$
and 
$$
\frm (p)=\sigma \theta(-p)\amdp e^{-\beta p^{2}/(2m)}\;,
$$
where $\sigma$ is an arbitrary positive constant (related to $\lambda$
in \eref{lambda}) and $\theta$ is the Heaviside function.
It is easy to verify, using Gaussian integration and the identity
$M=M\rho^{2}+m(1+\rho)^{2}$,
that the solution of equation (\ref{flfr}) is
given by
$$
g(P)=\sqrt{\frac{\beta }{2\pi M}}e^{-\beta
  P^{2}/(2M)}=\sqrt{\frac{\beta }{2\pi M}}e^{-\beta P^{2}(1-\rho
  )/((1+\rho )2m)}\;.
$$
Moreover, using the same identity several times one gets from
Eqs.\eref{out1} and \eref{out2} for the
exiting  distributions
$$
\flm (p)=\sigma \theta(-p)\amdp e^{-\beta p^{2}/(2m)}\;,
$$
and 
$$
\frp (p)=\sigma \theta(+p)\amdp e^{-\beta p^{2}/(2m)}\;.
$$
Therefore, we see that the Maxwellian fixed points (divided by $|p|$)
preserve both the
distribution $g$ of the scatterer, as well as the distributions of the
particles.

In fact, there are also non-Maxwellian fixed points of the form
$$
\flp (p)=\sigma \theta(+p)\amdp e^{-\beta  (p-m\flux)^{2}/(2m)}\;,
$$
and 
$$
\frm (p)=\sigma \theta(-p)\amdp e^{-\beta  (p-m\flux)^{2}/(2m)}\;.
$$
It is easy to verify that the solution of equation (\ref{flfr}) is now
given by
$$
g(P)=\sqrt{\frac{\beta }{2\pi M}}e^{-\beta  (P-M\flux)^{2}/(2M)}\;.
$$
Moreover, 
$$
\flm (p)=\sigma \theta(-p)\amdp e^{-\beta  (p-m\flux)^{2}/(2m)}\;,
$$
and 
$$
\frp (p)=\sigma \theta(+p)\amdp e^{-\beta  (p-m\flux)^{2}/(2m)}\;.
$$
The verification that this is a solution for any $a\in\real$ is again
by Gaussian integration. Note that if $a\ne0$ there is in fact a flux
through the cell.

\section{$N$ cells}

The model generalizes immediately to the case of $N$ cells which are
arranged in a row, by requiring that the exit distributions of any
given cell are equal to the entry distributions of the neighboring cells:
The cells are numbered from $1$ to $N$ and we have the collections of
functions $f^{+}_{{\rm L},i}$,  $f^{-}_{{\rm L},i}$, $f^{+}_{{\rm
    R},i}$, and  $f^{-}_{{\rm R},i}$ for the particle fluxes and $g_i$ for
the scatterers, $i=1,\dots,N$.
The equality of entrance and exit distributions is given by the identities  $f^{+}_{{\rm L},i+1}=f^{+}_{{\rm R},i}$, and 
$f^{-}_{{\rm R},i}=f^{-}_{{\rm L},i+1}$ for $1\le i<N$. 
The system is completely
determined by the two functions $f^{+}_{{\rm L},1}$ and $f^{-}_{{\rm
    R},N}$.
The equations \eref{flfr} 
generalize to 
\begin{equ}\label{flfrn}
g_i(P) =
\frac{1}{\rho\lambda}
\int_{\real} \d p \,g_i(\garg)\apdm\left (\flpp{i} (p)+\frmm{i}(p)\right)~,
\end{equ}
and similarly \eref{out1} and \eref{out2} lead to
\begin{equa}\label{flfrn2}
\atpdm \flmm{i} (\tilde p)&=\frac{\theta(-\tilde p)}{1-\rho }\int_{\real} \d p\,
g_i(\gargf)\,\apdm f_i (p)~,\\
\atpdm\frpp{i} (\tilde p)&=\frac{\theta(+\tilde p)}{1-\rho }\int_{\real} \d p\,
g_i(\gargf)\,\apdm f_i (p)\;,\\   
\end{equa}
where $f_i=\flpp{i}+\frmm{i}$.
Clearly, the Gaussians of the previous section are still solutions to
the full equations for $N$ contiguous cells.

Here we have closed the model by assuming independence between the
particles leaving and entering from the left (and from the
right). In concrete systems this is not true since a particle can
leave a cell to the left and re-bounce back into the original
cell after just one collision with
the scatterer in the neighboring cell, and, in such a situation there
is too much memory to allow for full independence.
It is possible to imagine several
experimental arrangements for which independence is a very good
approximation, see also \cite{EY2,EMZ2006} for discussions of such
issues. 
One of them could be to imagine  long channels between
the scatterers where time decorrelation would produce independence. Note
that ``chaotic'' channels may be  more complicated since they can modify
the distribution of left (right) traveling particles between two
cells.

\section{Continuous space}\label{s:continuous}

We are now ready to formulate the model in its final form. The cells
are now replaced by a continuum, with a variable $x\in[0,1]$ and the
relations we have derived so far will be generalized to this continuum
formulation. So we have moving particles, of mass $m$ and described by a
time-dependent density $f(t,p,x)$.

The scatterers have mass $M$ and their
momentum distribution is called $g(t,P,x)$. It is best to think that
the continuous variable $x\in[0,1]$ replaces the discrete index
$i\in\{0,\dots,N\}$. There is then an implicit rescaling of the form
$x\approx i/N$. Recall that the scatterers are \emph{fixed} in
space (although they have momentum) but that the particles will move
in the domain $[0,1]$.

We first impose, for all $x\in[0,1]$, the normalization
\begin{equ}\label{intg2}
\int_\real \d P\, g(t,P,x) = 1~,  
\end{equ}
which is the generalization of Eq.\eref{intg1}.
The particles again do not interact with each other, but only with the
scatterers and, expressed in momenta, the matrix maps
$(p,P)$ to $(\tilde p,\tilde P)$:
\begin{equ}\label{chocrel}
\left(\begin{matrix}
\tilde p\\
\tilde{P}
\end{matrix}\right)=
\left(\begin{array}{cc}
-\rho&(1-\rho)\\
(1+\rho)&\rho
\end{array}\right)\left(\begin{matrix}
p\\
P
\end{matrix}\right)\equiv S \left(\begin{matrix}
p\\
P
\end{matrix}\right)\;,
\end{equ}
as in Eq.\eref{Sv}.

We now rewrite the problem in the form of a Boltzmann equation, which
takes into account this matrix, as well as the particle
transport. One obtains, with $(\tilde p,\tilde P)$ related to $(p,P)$ as
above:
\begin{equa}\label{boltz1}
\partial_{t}f(t,p,x)&+\frac{p}{m}\partial_{x}f(t,p,x)
\\=&\int \d P\,
\left(\frac{|\tilde p|}{m} f(t,\tilde p,x)\,g(t,\tilde {P},x)
-\frac{| p|}{m} f(t,p,x)\,g(t,P,x)\right)~,\\
\partial_{t}g(t,P,x)=&\int\d p\,
\left(\frac{|\tilde p|}{m} f(t,\tilde p,x)\,g(t,\tilde {P},x)
-\frac{| p|}{m} f(t,p,x)\,g(t,P,x)\right)\;.\\
\end{equa}
The time independent version of the equation will be derived below
from the model with a chain of cells.
It is useful to introduce the function
$$
F(t,p,x)=\frac{|p|}{m}\,f(t,p,x)~,
$$
and then Eq.\eref{boltz1} takes the form
\begin{equa}\label{boltz2}
m\partial_{t}F(t,p,x)&+p\partial_{x}F(t,p,x)\\=&\,|p|\int\d P\,
\left(F(t,\tilde p,x)\,g(t,\tilde {P},x)
-F(t,p,x)\,g(t,P,x)\right)~,\\
\partial_{t}g(t,P,x)=&\int\d p\,
\left(F(t,\tilde p,x)\,g(t,\tilde {P},x)
-F(t,p,x)\,g(t,P,x)\right)\;.\\
\end{equa}

\begin{remark}
  One can also imitate a scattering cross section by introducing a
  factor $\gamma \in[0,1]$ in Eq.\eref{boltz2} (in the integrals) but
  this can be scaled away by a change of time and space scales. See
  also Sect.~\ref{dep}.
\end{remark}

We come now to the {\bf main equations} whose solutions will be discussed in
detail in the remainder of the paper.
The equation \eref{boltz2} takes, for the stationary solution, the
form
\minilab{eqstat}
\begin{equs}
\sign(p)\partial_{x}F(p,x)&=\int\d P\,
\left(F(\tilde p,x)\,g(\tilde {P},x)
-F(p,x)\,g(P,x)\right)~,\label{eqstat1}\\
0=&\int\d p\,
\left(F(\tilde p,x)\,g(\tilde {P},x)
-F(p,x)\,g(P,x)\right)\;.\label{eqstat2}
\end{equs}
We will show that this equation has non-equilibrium solutions.

\begin{remark}
  Note that the model we have obtained here is \emph{not}
  momentum-translation invariant, because of the term $\sign(p)$,
  except when the r.h.s.~of the equation is 0.
  \end{remark}

\noindent{\bf Derivation of \eref{eqstat}}. 
  The derivation of \eref{eqstat} from
  (\ref{flfrn}--\ref{flfrn2}) is based on the following formal limit:
We replace the index $i$ by the continuous variable $x=i/N$ and set
  $\epsilon =1/(2N)$. We consider that $f_{{\rm L},i}^{\pm}$ is at
  $(i-\HALF)/N=x-\epsilon $,
  while $f_{{\rm R},i}^{\pm}$ is at $x+\epsilon $.
We have the correspondences, with $\theta_\pm(p)\equiv\theta(\pm p)$:
\begin{equa}[2]
    \theta_+(p) F(p,x-\epsilon ) &=\apdm f_{{\rm L},i}^+~,\qquad
    \theta_-(p) F(p,x-\epsilon ) &=\apdm f_{{\rm L},i}^-~,\\
    \theta_+(p) F(p,x+\epsilon ) &=\apdm f_{{\rm R},i}^+~,\qquad
    \theta_-(p) F(p,x+\epsilon ) &=\apdm f_{{\rm R},i}^-~,\\
 g(P,x) &= g_i(P).
\end{equa}
To simplify momentarily the notation, let

\begin{equa}[2]
  F_-(p,x)\;\theta_+(p)  &\equiv \theta_+(p) F(p,x-\epsilon )~,\quad
  F_-(p,x)\;\theta_-(p)  &\equiv \theta_-(p) F(p,x-\epsilon )~,\\
  F_+(p,x)\;\theta_+(p)  &\equiv \theta_+(p) F(p,x+\epsilon )~,\quad
  F_+(p,x)\;\theta_-(p) &\equiv \theta_-(p) F(p,x+\epsilon )~.\\
\end{equa}
With these conventions, \eref{flfrn} becomes (setting $\lambda
  =1$):
 \begin{equ}\label{approxg}
   g(P,x) \,=\, \frac{1}{\rho} \int_\real \d q\, g({\textstyle\frac{P-(1+\rho) q}{\rho}},x)  \big(
   F_-(q,x)\theta_+(q) + F_+(q,x)\theta_-(q)\big)~,
 \end{equ}
which is equivalent to \eref{eqstat2}.
Similarly, Eq.\eref{flfrn2} leads to
\begin{equa}\label{10}
F_-(p,x)\theta_-(p)&=\frac{\theta(-p)}{1-\rho }\int_{\real} \d q\,
g({\textstyle{\frac{p+\rho q}{1-\rho
}}},x)\,\big(F_-(q,x)\theta_+(q)+F_+(q,x)\theta_-(q)\big)~,\\
F_+(p,x)\theta_+(p) &=\frac{\theta(+ p)}{1-\rho }\int_{\real} \d q\,
g({\textstyle{\frac{p+\rho q}{1-\rho
}}},x)\,\big(F_-(q,x)\theta_+(q)+F_+(q,x)\theta_-(q)\big)\;.\\    
\end{equa}
Subtracting the first equation from the second in \eref{10} leads to
\begin{equa}\label{part1}
  F_+(p,x)\theta_+(p)&- F_-(p,x)\theta_-(p)\\\,=\,
\frac{\sign(p)}{1-\rho }& \int _\real \d q\, 
g({\textstyle{\frac{p+\rho q}{1-\rho }}},x) \
\big(F_-(q,x)\theta_+(q)+F_+(q,x)\theta_-(q)\big)~.
\end{equa}
On the other hand,  since $\d p\, \d P=\d\tilde p\, \d\tilde{P}$ we
can, by \eref{eqstat2}, 
impose the condition
$$
\int \d P\, g(P,x)=1~,
$$
for all $x$. Then we have the trivial identity
\begin{equ}\label{part2}
  F_-(p,x)\theta_+( p) - F_+(p,x)\theta_-(p)
 =
\int _\real \d P\, 
g(P,x)  \big(F_-(p,x)\theta_+( p) - F_+(p,x)\theta_-(p)\big)~.
\end{equ}
Subtracting \eref{part2} from \eref{part1}, we get for $p>0$,
\begin{equa}\label{part3}
F_+(p,x)-F_-(p,x)\,=\,  
  &\frac{1}{1-\rho }\int \d q\,g({\textstyle{\frac{p+\rho q}{1-\rho
}}},x)\big(F_-(q,x)\theta_+(q)+F_+(q,x)\theta_-(q)\big)\\&- \int \d q\, g(q,x)
F_-(p,x)
\end{equa}
while for $p<0$,
\begin{equa}\label{part4}
F_+(p,x)-F_-(p,x)\,=\,   
 -&\frac{1}{1-\rho }\int \d q\,g({\textstyle{\frac{p+\rho q}{1-\rho}
}},x)\big(F_-(q,x)\theta_+(q)+F_+(q,x)\theta_-(q)\big)\\&+ \int \d q\, g(q,x)
F_+(p,x)
\end{equa}
Note now that
\begin{equa}
  F_-(q,x)\theta_+(q)+&F_+(q,x)\theta_-(q)\\
&=F(q,x-\epsilon )\theta_+(q)+F(q,x+\epsilon )\theta_-(q)\\
&=F(q,x)+\theta_+(q)\bigl(F(q,x-\epsilon )-F(q,x)\bigr)\\
&\hphantom{=F(q,x)~}+\theta_-(q)\bigl(F(q,x+\epsilon )-F(q,x)\bigr)~.
\end{equa}
Therefore, replacing the $F_\pm$ in the r.h.s.~in \eref{part3} and \eref{part4} by
$F(p,x)$ is a higher order correction in $\epsilon $, and we
finally get
\begin{equ}
  F(p,x+\epsilon )-F(p,x-\epsilon )=\frac{\sign(p)}{1-\rho }
\int_\real \d q\bigg(\,g({\textstyle{\frac{p+\rho q}{1-\rho}
}},x) F(q,x) - g(q,x) F(p,x)
\bigg)~.
\end{equ}
A further change of integration variables leads to \eref{eqstat1},
while \eref{approxg} leads to \eref{eqstat2}. (We have not taken into
account the scaling by $\epsilon=1/(2N) $ which is needed to get the
derivative; we will come back to this question in the discussion in Sect.~\ref{dep}.) This ends the derivation of \eref{eqstat}.

The derivative term in Eq.\eref{eqstat} reflects the gradients which
have to appear when the system is out of equilibrium.
However, if the system is at equilibrium, the equivalence between
Eq.\eref{eqstat} and Eqs.\eref{flfr}--\eref{out2} immediately tells us
that stationary solutions in the form of Gaussians (for $F$, not for
$f$) exist:
\begin{equ}\label{stat}
  F(p)=\gamma\sqrt{\frac{\beta }{2\pi m}}e^{-\beta p^2/(2m)}~,\qquad g(P)=\sqrt{\frac{\beta }{2\pi M}}
  e^{-\beta P^2/(2M)}~.
\end{equ}
Furthermore, we have again translated versions of this fixed point,
\begin{equ}\label{stata}
  F(p)=\gamma\sqrt{\frac{\beta }{2\pi m}}e^{-\beta (p-ma)^2/(2m)}~,\qquad g(P)=\sqrt{\frac{\beta }{2\pi M}}
  e^{-\beta (P-Ma)^2/(2M)}~,
\end{equ}
because in this case, the r.h.s.~of Eq.\eref{eqstat} is zero.

\subsection{Flux}\label{s:flux}

We can define various fluxes of the particles (recall that
$F(p,x)=|p|f(p,x)/m$):
\begin{equa}[3]\label{fluxes}
\Phi_{\rm P}&=& \text{particle flux}&=&\int\d p\,&\sign(p)\,F(p,x)\;, \\  
\Phi_{\rm M}&=&~ \text{momentum activity }&=&\int\d p\,& |p|\,F(p,x)\;, \\
\Phi_{\rm E}&=& \text{energy flux}&=&\int\d p\,&\frac{\sign(p) p^{2}}{2m}\,F(p,x)\;. \\
\end{equa}
Note that for the stationary Maxwellians of \eref{stata} these fluxes
are equal to
\begin{equa}
 \Phi_{\rm P}=&\sqrt{\frac{2m\pi}{\beta}}\,\,\text {erf}({a\sqrt{\beta m/2}})~,\\
\Phi_{\rm M}=&\frac{2m}{\beta }e^{-\beta ma^2/2} +a 
\sqrt{\frac{2 m^3\pi}{\beta }}\text{erf}( {a\sqrt{\beta m/2}})~,\\
\Phi_{\rm E}=&\frac{2am^2}{\beta } e^{-\beta ma^2/2}+(1+\beta
ma^2)\sqrt{\frac{2 m^3\pi}{\beta ^3}} \text{erf}( {a\sqrt{\beta m/2}})~.\\
\end{equa}

Also note that for $a=0$ the quantity $\Phi_{\rm M}$ does not
vanish. This is because it measures the total outgoing flux, not the
directed outgoing flux (which is of course 0 when $a=0$).

\begin{lemma}
For every stationary solution of \eref{eqstat} the 3 fluxes of
\eref{fluxes} are independent of $x\in[0,1]$.
\end{lemma}

\begin{proof}
From \eref{eqstat1}  we deduce that
$$
\partial_{x} \int\d p\,\sign(p)\,F(p,x)=
\int\d p\, \d P\,\left(F(\tilde p,x)\,g(\tilde {P},x)
-F(p,x)\,g(P,x)\right)\;,
$$
which vanishes since $\d p\, \d P=\d\tilde p\, \d\tilde P $.
Similarly, multiplying \eref{eqstat1} by $p$ and
integrating over $p$, we get
$$
\partial_{x} \int \d p\,|p|\,F(p,x)=
\int\d p \,\d P\,\,\, p\left(F(\tilde p,x)\,g(\tilde {P},x)
-F(p,x)\,g(P,x)\right)\;.
$$
Multiplying \eref{eqstat2} by $P$ and
integrating over $P$, we get
$$
0=\int \d p\,\d P\,\, P
\left(F(\tilde p,x)\,g(\tilde {P},x)
-F(p,x)\,g(P,x)\right)\;.
$$
Adding these two equations, we see that
$$
\partial_{x} \int \d p\,|p|\,F(p,x)=
\int\d p\,\d P\,\, (p+P)\left(F(\tilde p,x)\,g(\tilde {P},x)
-F(p,x)\,g(P,x)\right)\;.
$$
But this vanishes, since $P+p=\tilde P +\tilde p$ by momentum
conservation, and using again $\d p\, \d P=\d\tilde p\, \d\tilde P $.
In a similar way, we first have
\begin{equa}
\partial_{x} &\int\d p\, \frac{|p|p}{2m}\,F(p,x)\\&=
\int\d p\, \d P\,\, \frac{p^{2}}{2m}\left(F(\tilde p,x)\,g(\tilde {P},x)
-F(p,x)\,g(P,x)\right)\\
\end{equa}
Finally, multiplying this time \eref{eqstat2} by $P^2/M$, 
integrating over $P$, adding to the above equation and using energy
conservation, we get
\begin{equa}
\partial_{x} &\int\d p\, \frac{|p|p}{2m}\,F(p,x)\\&=
\int\d p\, \d P\,\, \left(\frac{p^{2}}{2m}+\frac{{P}^{2}}{2M}\right)\;
\left(F(\tilde p,x)\,g(\tilde {P},x)
-F(p,x)\,g(P,x)\right)\\&=0\;.\\
\end{equa}
Thus, all three fluxes are independent of $x$, as asserted.
\end{proof}

\section{Formulating the heat-conduction problem}

Based on the stationarity equation \eref{eqstat}, we now formulate the
problem of heat conduction in mathematical terms. We imagine that the
input of the problem is given by prescribing the \emph{incoming} fluxes on
both sides of the system. The system, in its stationary state, should
then adapt all the other quantities, $F$ and $g$, to this given input,
which describes really the forcing of the system. In particular, the
distribution of the outgoing fluxes will be entirely determined by the
incoming fluxes.

We now formulate this question in mathematical terms:  
The incoming fluxes are described by two functions
$\mathscr{F}_{0}(p)$ (defined for  $p\ge0$) and  
$\mathscr{F}_{1}(p)$ (defined for $p\le0$). These are the incoming
distributions on the left end (index 0) and the right end (index 1) of
the system.

In terms of these 2 functions, the problem of existence of a
stationary state can be formulated as (recall that the rescaled system has length one):

{\it 
Is there a solution $(F,g)$ of the equations \eref{eqstat}
with the boundary conditions
\begin{equ}\label{boundary}
F(p,0)=\mathscr{F}_{0}(p)~,\;\,\forall p\ge 0
\qquad \text{and}\qquad  F(p,1)=\mathscr{F}_{1}(p)~,\;\,\forall p\le 0\;.
\end{equ}
}

Assume for a moment that, instead of the boundary conditions
\eref{boundary} we were given just $F(p,0)$, but now \emph{for all
  $p\in\real$}, not only for $p>0$. Assume furthermore, that $g(p,x)$
is determined by \eref{eqstat2}.
In that case, the relation \eref{eqstat} can be written as a dynamical
system in the variable $x$:
\begin{equ}\label{evolution}
\partial_{x}F(\cdot,x)=\mathscr{X}(F(\cdot,x))~.  
\end{equ}
Thus, if $F(\cdot,0)$ is given, then, in principle, $F(\cdot,1)$ is
determined (uniquely) by Eq.\eref{evolution}, provided such a solution
exists. We denote this map by
$\YY_0$: 
$$
\YY_0: F(\cdot,0) \mapsto  F(\cdot,1)~.
$$
What is of interest for our problem is the restriction of the image of
$\mathscr{Y_0}$ to functions of negative $p$ only, since that
corresponds to the incoming particles from the right side, and so we define
$$
\biggl(\YY(F(\cdot,0)) \biggr)(p) \equiv \theta_-(p)\,\cdot\,\biggl(\YY_0( F(\cdot,0))\biggr)(p)=\theta_-(p)\cdot F(p,1)~.
$$

Using this map $\YY$,
we will show that
when $F(\cdot,0)$ varies in a small neighborhood
 the map $\YY$ is invertible on its image. 
By taking inverses the
problem of heat conduction for our model will be solved for small
temperature and flux difference.

Of course, this needs a careful study of the function space on which
$\YY$ is supposed to act. This will be done below.

To formulate the problem more precisely, we change notation, and let 
\begin{equa}
  F_0^+(p)&=\theta_+(p)F(p,0)~,\\
  F_0^-(p)&=\theta_-(p)F(p,0)~,\\
  F_1^+(p)&=\theta_+(p)F(p,1)~,\\
  F_1^-(p)&=\theta_-(p)F(p,1)~.\\
\end{equa}
We assume now that $F_0^+$ is \emph{fixed} once and for all and omit it
from the notation.
Then, we see that $\YY$ can be interpreted as a map which maps the
function $F_0^-$ to $F_1^-$, and we call this map $\Phi$. 

We will show below that for $F_0^-$ in a small
neighborhood $\domain$  the map $\Phi$ is 1-1 onto its image
$\Phi(\domain)$ and can therefore be inverted. For any $\hat F$ in
$\Phi(\domain)$, we can
take $F_0^-=\Phi^{-1} (\hat F)$, and
we will have solved the problem of existence of heat flux. 

\section{The $g$ equation}\label{s:gequ}

We start here with the study of existence of $g$ for given $F$. Since 
\eref{eqstat2} does not couple different $x$, we can fix $x$.
The equation \eref{eqstat2} is then equivalent to 
$$
g=\mathcal{A}_{F}(g)~,
$$
where the operator $\mathcal{A}_{F}$ acting on the function $h$ is defined by
$$
\mathcal{A}_{F}(h)(\tilde P,x)=\frac{\int \d p\,
F\big(\tilde p(p,\tilde {P}),x\big)h\big(P(p,\tilde {P}),x\big)\,}{\int\d p\,
 F\big(p,x\big)}\;,
$$
(provided the
denominator does not vanish).
Note that for fixed  $\tilde {P}$ and $p$, we can solve the collision
system \eref{chocrel} to find the corresponding $P$ and $\tilde p$,
namely
$$
P=\frac{1}{\rho}\;\tilde {P}-\frac{1+\rho}{\rho}\;p\qquad\text{and}
\qquad \tilde p=\frac{1-\rho}{\rho}\;\tilde {P}-\frac{1}{\rho}\;p\;.
$$
The action of the operator $\mathcal{A}_{F}$ can then be rewritten as
\begin{equ}\label{legop}
\mathcal{A}_{F}(h)({\tilde P},x)=\frac{\rho ^{-1}\int\d p\,
F(p,x)h\big(\tilde {P}/\rho-(1+\rho)p/\rho,x\big)}{\int\d s\,
 F(s,x)}\;.
\end{equ}

In order to study this operator notice that it does not depend
 explicitly on $x$.  It is convenient to study instead a 
family of operators indexed by
functions $\phi$ of the momentum only. We define (assuming the
integral of $\phi$ does not vanish)
\begin{equ}
\big(\ope_{\phi}\psi\big)(\tilde P)=
\frac{\int\d p\,
\phi(p)\psi\big(\tilde P/\rho-(1+\rho)p/\rho\big)}{\rho\int
\d p\,\phi(p)}\;.
\end{equ}

A final change of variables will be useful when we study $\ope_\phi$:
\begin{equ}\label{opg}
\big(\ope_{\phi}\psi\big)(p)=
\frac{\int\d q\,
\phi(\varg)\psi\big(q)}{(1+\rho )\int
\d q\,\phi(q)}\;.
\end{equ}

\section{The mathematical setup and the main result}\label{s:noch}

Having formulated the problem of existence of the stationary solution
in general, we now fix the mathematical framework in which we can
prove this existence. This framework, while quite general, depends
nevertheless on a certain number of technical assumptions which we
formulate now.

We fix once and for all the ratio $\mu=m/M$ of the masses, and assume,
for definiteness, that $\mu\in (0,1)$. It seems that this condition is
not really necessary, and probably the condition $m\ne M$ (and the
masses non-zero) should work as well, but we have not pursued this.

We next describe a condition on the incoming distribution, called $F$
in the earlier sections. The basic idea, inspired from the equilibrium
calculations, is that $F(p,x)$ should be close to
\begin{equ}
  F_{\rm reference}(p)=
    \exp(-\beta p^2/(2m))\equiv \exp(-\alpha p^2)~,
\end{equ}
while the derived quantity $g(p,x)$ should be close to
\begin{equ}
  g_{\rm reference}(p)=
    \exp(-\beta  p^2/(2M))\equiv \exp(-\mu \alpha  p^2)~.
\end{equ}
Upon rescaling $p$, we may assume henceforth that $\alpha =1$.

The operators of the earlier sections will now be described in spaces
with weights $$
\WW_\nu(p)=\exp(-\nu\ap)~,
$$
where we will choose $\nu=1$ for the $F$ and $\nu=\mu$ for the $g$.

We recall that the operator $\ope_F$ in ``flat'' space is
$$
(\ope_Fg)(p)\,=\,\frac{\int \d q\, F(\varg) g(q)}{(1+\rho )\int \d q F(q)}~.
$$
We then define the
integral kernel in the space with weights $\exp(-\ap)$ for $F$ and
$\exp(-\mu\ap)$ for $g$, and write
\begin{equa}
  F(p)\,=\,e^{-\ap} v(p)~,\qquad 
    g(p)\,=\,e^{-\mu\ap} u(p)~. 
\end{equa}
Here, $\mu=m/M=(1-\rho )/(1+\rho )$, as before.
Expressed with $u$ and $v$ the operator $\ope_F$ takes
the form
\begin{equ}\label{luv}
  \big(K_{v}u \big)(p) = \frac{1}{(1+\rho )\int \d q e^{-\aq} v(q)}\cdot \big(L_v u\big)(p)~,
\end{equ}
where
\begin{equ}
  (L_v u)(p)\,=\,\int \d q\, v(\varg) K(p,q) u(q)~,
\end{equ}
and
\begin{equ}
  K(p,q)= \WW_1({\textstyle{\frac{p-\rho q}{1+\rho
  }}})\cdot\WW_\mu(q)/\WW_\mu(p) ~.
\end{equ}
A simple calculation shows that
\begin{equ}\label{simplekpq}
K(p,q)\,=\,e^{- (\rho p-q)^2/(1+\rho )^2}~.
\end{equ}
Our task will be to understand under which conditions the linear operator
$\ope_F$ has an eigenvalue 1. This will be done by showing that $K_v$
is quasi-compact. It is here that we were not able to give reasonable
bounds on $K(p,q)$ in the case of different exponentials for $p>0$ and
$p<0$, which represents different temperatures for ingoing and outgoing particles.

\subsection{Function spaces}

We now define spaces which are adapted to the simultaneous requirement
of functions being close to a Gaussian near $|p|=\infty $ and $u$ and
$v$ having 
limits, \emph{and} $K_v$ being quasi-compact.
We define a space $\banach_1$ of functions $u$ with norm
$$
\|u\|_{\banach_1}=\int e^{-\mu \ap}|u(p)|\d p~.
$$
Similarly, $\vbanach_{1}$ is the space of functions $v$
with norm
$$
\|v\|_{\vbanach_1}=\int e^{- \ap}|v(p)|\d p~.
$$
Thus, the only difference is the absence of the factor $\mu=(1-\rho
)/(1+\rho )$ in the exponent.

We also define a smaller space
$\banach_{2}$, contained in  $\banach_{1}$, with the norm
$$
\|u\|_{\banach_{2}}=\int |\d u(p)|+ \int
e^{-\mu \ap}|u(p)|\d p\;,
$$
and the analogous space $\vbanach_{2}$ contained in  $\vbanach_{1}$
with the norm 
$$
\|v\|_{\vbanach_{2}}=\int |\d v(p)|+ \int
e^{- \ap}|v(p)|\d p\;,
$$

\begin{remark}
To simplify notation we write $\int |\d u(p)|$ instead of the
variation norm. However, the ``integration by parts'' formula would
hold with the ``correct'' definition of variation as well.
\end{remark}

\begin{lemma}\label{lepremier}
One has the inclusion $\banach_{2}\subset {\rm L}^{\infty}$, and more
  precisely
$$
\|u\|_{{\rm L}^{\infty}}\le\int |\d u|+e^{\mu} \int e^{-\mu \ap}
|u(p)|\d p\,\le\,
e^\mu\|u\|_{\banach_2}\;. 
$$
Furthermore, if $u\in \banach_{2}$, then
  $\lim_{p\to\pm\infty}u(p)$ exists.  The maps $u\mapsto \lim_{p\to\pm\infty }
  u(p)$ and
$u\mapsto \int \d p\, \exp(-\mu \ap)\cdot u(p) $ are continuous functions
  from $\banach_2$ to $\real$. The unit ball of $\banach_{2}$ is
  compact in $\banach_{1}$.

Analogous statements hold for the spaces $\vbanach_2$ (defined without
the factor $\mu$).
\end{lemma}

\begin{proof}
The first  statement is  easy, but it will be convenient to have
the explicit  estimates. We have
$$
u(y)-u(x)=\int_{[x,y]}\d u~,
$$
and therefore
$$
|u(x)|\le \int |\d u|+\int_{-1/2}^{1/2}|u(y)|\d y\le 
\int |\d u|+e^{\mu} \int e^{-\mu \ap} |u(p)|\d p\;.
$$
The second statement follows at once since the
functions in $\banach_{2}$ are of bounded variation. 

For the last assertions, it follows from the inclusion in ${\rm L}^{\infty}$ 
that the unit ball of $\banach_{2}$ is equi-integrable at infinity in
${\rm L}^{1}\big(e^{-\mu \ap}\d p\big)$. Moreover, a set of uniformly
bounded 
functions
of uniformly bounded variation is
compact in any ${\rm L}^{1}(K,\d p)$ for any compact subset $K$ of $\real$ (see
\cite{DSII}, Helly's selection principle). 
\end{proof}

\subsection{A cone in $\vbanach_2$}

We will work in the space $\vbanach_2$ but we will need a cone (of
positive functions, with adequate decay) in this space, in order to
prove quasi-compactness of $K_v$.

We define 
a cone $\mathcal{C_\vbanach}$ in $\vbanach_{2}$ by the condition
\begin{equ}\label{e:thecone}
\mathcal{C_\vbanach}=\left\{ v\in\vbanach_2~,~~v\ge
0\text{ and }Z\cdot\lim_{p\to\pm\infty}v(p)<1\right\}~, 
\end{equ}
where 
\begin{equ}\label{Z}
  Z=Z(v)=\frac{\sqrt{\pi}}{\int \d p \,\,e^{-\ap} v(p)}~.
\end{equ}
\begin{lemma}
The cone $\mathcal{C_\vbanach}$ has non empty interior (in
$\vbanach_{2}$) and is convex.
\end{lemma}
\begin{proof}
By Lemma \ref{lepremier} the maps $v\mapsto \lim_{p\to\pm\infty}v(p)$ and $v\mapsto\int
\exp(-\ap)v(p) \d p $ are continuous in $\vbanach_{2}$ and hence the
assertion follows.
\end{proof}
\begin{remark}
Note that a function in the interior of the cone is necessarily
bounded away from zero, since at infinity it must have a non-zero limit
and in any compact set, if it is never zero, it is bounded away from zero.
\end{remark}

\begin{remark}
  Note that the function $v\equiv 1$
  (the Gaussian) is \emph{not} in the cone $\vcone$. In fact, we
  require that $\lim_{p\to\pm\infty} F(p)e^{p^2}\cdot \int e^{{-p'}^2}\d p'/\int F(p')\d p'< 1$.
\end{remark}

\subsection{The main result}

On the set $\vcone$, we consider now the spatial evolution
equations \eref{eqstat} in the variables $v $ and $u_v $ (which
is the solution of $K_{v}u =u$ with $K_v$ defined in \eref{luv}):
\begin{equa}\label{champ}
\partial_{x}&v (p,x)\\&=\sign(p)\int\d P\,
\bigg((\WW_1\kern-0.2em\cdot\kern-0.2em v) (\tilde p,x)\,\,
(\WW_\mu\kern-0.2em\cdot\kern-0.2em u_{v (\cdot,x)})(\tilde {P},x)\\
&~~~~~~~~~~~~~~~~~~~~~~~~~~~~~~~~-(\WW_1\kern-0.2em\cdot\kern-0.2em v) (p,x)\,\,
(\WW_\mu\kern-0.2em\cdot\kern-0.2em u_{v (\cdot,x)})(P,x)\bigg)\\
&=
\sign(p)\int\d P\,
(\WW_1\kern-0.2em\cdot\kern-0.2em v) \big(-\rho p+(1-\rho) P,x\big)\,
(\WW_\mu\kern-0.2em\cdot\kern-0.2em u_{v (\cdot,x)})\big((1+\rho)p+\rho P,x\big)\\
&\ \ \ -\sign(p)\,\,v (p,x)\int\d P\, e^{-\mu P^2}u_{v (\cdot,x)}(P,x)~,\\
\end{equa}
with initial condition $v (\cdot,x=0)\in\vcone$. We will give a more
explicit variant in \eref{lav}.

Any solution of this equation is a function of $p$ and $x$, and it is
easy to verify that it satisfies  the equation
\eref{eqstat1}. Together with the definition of $u_{v }$ we have a
complete solution of the nonlinear system (\ref{eqstat}). Here we assume
of course that the r.h.s.~of the above equation is well defined as a
function, so that we can multiply by $\sign(p)$.

\begin{theorem}\label{thm:main}
For any $v _{0}\in \vcone$, there are a number $x_{v _{0}}>0$ 
and a neighborhood $\voisin_{v _{0}}$ of $v _{0}$ in
$\vcone$ such that the 
solution  of (\ref{champ}) exists for any initial condition
$v_0=v(p,0)\in\voisin_{v _{0}}$ and for any $x$
in  the interval $[0,x_{v _{0}}]$. The function $v _0\mapsto x_{v _0}$ is
continuous from  
$\vcone$ to $\real^{+}$ compactified at infinity. We denote
by $\Phi_x $ the semi-flow integrating (\ref{champ}). For any
$x\in[0,x_{v _{0}}]$, the map
$\Phi_x:v\mapsto \Phi_x(v)$ is a local diffeomorphism, \ie, a diffeomorphism on
$\voisin_{v _{0}}$.  
\end{theorem}

Note that this implies in particular that the probability densities
for $g(\cdot,x)$ and $F(\cdot,x)$ remain positive for all
$x\in[0,x_{v_0}]$,  which is of course crucial from the physics point
of view.

We will prove this in Sect.~\ref{f1} (and in the appendix).

\section{Bound on the operator $K_v$ and proof of Theorem~\ref{thm:main}}\label{f1}

These bounds are the crux of the matter. They actually show, that,
under the conditions on $\vbanach_2$ and the set $\vcone$, the
operator $K_v$ is quasi-compact. In terms of the physical problem, this
means that the scatterer is not heating up if the incoming fluxes are
in $\vbanach_2$.

The object of study is, for $v\in \vcone$, the operator
$$
\big(K_{v}u\big)(p)=\frac{1}{(1+\rho)\int e^{-\aq}v(q)\d q}
\int v\big(\varg \big)K(p,q) u(q)\d q\;. 
$$
and we are asking for a solution $u$ of the equation
$K_{v}(u)=u$.

\begin{lemma}
If $v\ge 0$, $K_{v}$ is a positive (nonnegative) operator, and
$$
\int \d p\,e^{-\mu \ap}\big(K_{v}u\big)(p) =
\int \d p\,e^{-\mu \ap}u(p)
$$
and
$$
\big\|K_{v}\big\|_{\banach_{1}}=1\;.
$$
\end{lemma}
\begin{proof}
Easy, compute and take absolute values. Alternately, consider that the
probability is conserved in the original space.
\end{proof}

Our main technical result is

\begin{proposition}\label{main}
For $v\in\vcone$, there exist a $\zeta $, $0\le \zeta<1 $ and an $R>0$ (both depend on
$v$ continuously) such that for any $u\in\banach_{2}$ one has the bound
$$
\int \big|\d K_{v}(u)\big|\le \zeta \int \big|\d u\big|+R
\|u\|_{\banach_{1}}\;. 
$$
\end{proposition}

\begin{proof}

Since $v$ will be fixed throughout the study of $K_v$, it will be
useful to introduce the abbreviation $Q=Q_v$ for the normalizing
factor
\begin{equ}\label{Q}
  Q=\frac{1}{(1+\rho)\int e^{-\aq}v(q)\d q}~.
\end{equ}

We will use a family of smooth cut-off functions $\chi_{L}$
($L> 1$) which are equal to 1 on $[-L+\HALF,L-\HALF]$ and which vanish on $|q|>L+\HALF$. Let $\Theta$ be a
$C^{\infty}$  function satisfying $0\le \Theta\le 1$, with
$\Theta(q)=0$ for $q\le -\HALF $ and $\Theta(q)=1$ for $q\ge \HALF $.
We define $\chi_{L}$ by
$$
\chi_{L}(q)=
\begin{cases}
\Theta(q+L)\;&\text{if}\;  q\le -L+\HALF \;,\\
1\;&\text{if}\; -L+\HALF \le q\le L-\HALF \;,\\
\Theta(L-q)\;&\text{if}\; q\ge L-\HALF \;.\\
\end{cases}
$$
The functions $\chi_{L}$ are $C^{\infty}$, satisfy $0\le\chi_{L}\le 1 $
and $\big\|\chi_{L}' \big\|_{{\rm L}^{\infty}}$ is independent of $L$.
Let $L_{1}$ and $L_{2}$ be two positive numbers  to be chosen large
enough later on (depending on $v$).
We will use the partition of unity
$$
1=\chi_L+\chi_L^\perp~.
$$
Using  this decomposition of unity with $L=L_{1}$ and $L=L_{2}$, we write
$K_{v}=K_{v}^{(1)}+K_{v}^{(2)}+K_{v}^{(3)}$ with
\begin{equa}
\big(K_{v}^{(1)}u\big)(p)
&=Q
\int \d q\,v\big(\varg \big)K(p,q)u(q)\cdot\chi_{L_{1}}(q)\;,\\
\big(K_{v}^{(2)}u\big)(p)&=Q\int \d q\,  
  v\big(\varg \big)K(p,q)u(q)\cdot\chi^\perp_{L_{1}}(q)\,\chi_{L_{2}}(p-\rho q)
\;,\\
\big(K_{v}^{(3)}u\big)(p)&=Q
\int \d q\,
  v\big(\varg \big)K(p,q)u(q)\cdot\chi^\perp_{L_{1}}(q)\,\chi^\perp_{L_{2}}(p-\rho q)\,
\;.
\end{equa}
We will now estimate the variation of the three operators separately.

For the variation of the first term, we find
\begin{equa}
\d\big(K_{v}^{(1)}u\big)(p)
=&+Q
\int \d q\, \d v\big(\varg \big){\textstyle\frac{1}{1+\rho }}\cdot K(p,q) u(q) \chi_{L_{1}}(q)\\
&+Q\,\d p
\int \d q\,v\big(\varg \big)\partial_p K(p,q)\cdot u(q)\chi_{L_{1}}(q)~.\\
\end{equa}
Using the explicit form of $K(p,q)$  (see Eq.\eref{simplekpq}), and
some easy bounds which we defer to the Appendix,
we get the bound
\begin{equa}\label{e:summaryk1}
\int& \big|\d\big(K_{v}^{(1)}u\big)(p)\big|\le   
\Oun Q
\left(\|v\|_{{\rm L}^{\infty}}+\int |\d
v|\right)\int_{-L_1-\HALF }^{L_1+\HALF }|u(q)|\d q\\
&\le 
\Oun Q
\left(\|v\|_{{\rm L}^{\infty}}+\int |\d v|\right) e^{\mu(L_1+\HALF )^{2}} 
\int e^{-\mu\aq}|u(q)|\d q\,\\
&\le\,\const \|u\|_{\banach_1}\cdot\|v\|_{\vbanach_2}\;.
\end{equa}

The variation of $K_v^{(2)}$ leads to three terms:
\begin{equa}
\d\big(K_{v}^{(2)}u\big)&(p)\\
=&\,Q\,\d p
\int \d q\,
  v\big(\varg \big)K(p,q) u(q)\cdot\chi^\perp_{L_{1}}(q)\,\chi_{L_{2}}'(p-\rho q)
\\
+ &\,Q\,
\int \d q\,
 \d v\big(\varg \big){\textstyle\frac{1}{1+\rho }}\cdot K(p,q) u(q)\cdot \chi^\perp_{L_{1}}(q)\, \chi_{L_{2}}(p-\rho q)
\\
+& Q\,\d p\,
\int \d q\,
  v\big(\varg \big)\;\partial_pK(p,q)\,\cdot u(q)\cdot\chi^\perp_{L_{1}}(q)\,
\chi_{L_{2}}(p-\rho q)\\
:&=\d J_{21}+\d J_{22}+\d J_{23}\;.
\end{equa}
In these terms, the variables $p$ and $q$ are in the domain 
$$
\DD=\{ (p,q)\in \real^2~:~ |p-\rho q|< L_2+\HALF\text{ and } |q|
    > L_1-\HALF\}~,
$$
and for
$L_1=3\rho L_2/(1-\rho ^2)$ and $L_2$ sufficiently large we have from Lemma~\ref{l:kpq}:
\begin{equ}\label{boundonk}
  K(p,q)< \exp\left(-C_1 (\rho
  p-q)^2-C_2 L_2^2\right)~.
\end{equ}
Therefore, we get for $\d J_{21}$:
\begin{equ}\label{e:J1}
  \int |\d J_{21}| \le \const e^{-C_2L_2^2} \|u\|_\infty \|v\|_\infty
  \int_{(p,q)\in\DD} \d p\, \d q e^{-C_1 (\rho
  p-q)^2}~.
\end{equ}
The integral exists and is uniformly bounded in $L_2$ (since $|\rho
p-q|\to \infty $ when $|q|\to\infty $).

The term $\d J_{22}$ is handled in a similar way and leads to the bound
\begin{equ}\label{e:J2}
  \int |\d J_{22}| \le \const e^{-C_2L_2^2} \|u\|_\infty \int |\d v|~.
\end{equ}

The following identity is useful:
\begin{equ}\label{truc2}\label{truc}
\partial_p K(p,q) = -\rho \partial_q K(p,q)~.
\end{equ}

For the term $\d J_{23}$ we observe that from \eref{truc} one gets,
upon integrating by parts, with the notation 
\begin{equa}
\d J_{23}
=&\rho  Q\, \d p
\int \d q\,
v\big(\varg
\big)K(p,q)\cdot\partial_q\left(\chi^\perp_{L_{1}}(q)\chi_{L_{2}}(p-\rho
q) \right)  \\
&+\rho   Q\,\d p
\int 
  \d v\big(\varg \big){\textstyle\frac{-\rho }{1+\rho }}K(p,q)u(q)\cdot\chi^\perp_{L_{1}}(q)\chi_{L_{2}}(p-\rho
q) 
\\
&+\rho Q\,\d p
\int 
  v\big(\varg \big)K(p,q)
\d u(q)\cdot\chi^\perp_{L_{1}}(q)\chi_{L_{2}}(p-\rho
q) \\
:=& \, \d J_{231} + \d J_{232} + \d J_{233}~.
\end{equa}
All these terms are localized in the domain $\DD$. In $\d J_{231}$ 
there appears a derivative 
\begin{equa}
X=&\partial_q\left(\chi^\perp_{L_{1}}(q)\chi_{L_{2}}(p-\rho
q) \right)\\
=&-\chi'_{L_{1}}(q)\chi_{L_{2}}(p-\rho
q) \\
&-\rho \chi^\perp_{L_{1}}(q)\chi'_{L_{2}}(p-\rho
q) \\
:=& X_1+X_2~.
\end{equa}
The terms involving $X_1$ and $X_2$ can be bounded as $\d J_{21}$ and $\d
J_{22}$ by observing that $\supp \chi'_{L_1}  \subset \{ |q| <
L_1+\HALF\}$, and similarly for $X_{2}$.

The terms $\d J_{232}$ and $\d J_{233}$ are bounded similarly.

Together, these lead to a bound
\begin{equ}\label{e:summaryk2}
\int \big|\d K_v^{(2)} (u)\big|\le \const e^{-C_2L_2^2} \|
v\|_{\vbanach_2}\|u\| _{\banach_2}~.
\end{equ}
\begin{remark}
  Note that in this term, the norm $\|u\| _{\banach_2}$ appears with a
  \emph{small} coefficient, while in \eref{e:summaryk1} it was $\|u\|
  _{\banach_1}$ (with a large coefficient).
\end{remark}

Finally, we estimate the total variation of
$K^{(3)}_{v}(u)$ and here, the nature of the set $\vcone$ will be important.
We have
\begin{equa}
\d\big(K_{v}^{(3)}&u\big)(p)\\=&Q
\int \d q\,
 \d v\big(\varg \big){\textstyle\frac{1}{1+\rho }}K(p,q)u(q)\cdot\chi^\perp_{L_{1}}(q)\, \chi^\perp_{L_{2}}(p-\rho q)\,
\\
-&Q\,\d p  
\int \d q\,v\big(\varg \big)K(p,q)u(q)\cdot\chi^\perp_{L_{1}}(q)\,
\chi_{L_{2}}'(p-\rho q)
\\
+&Q\,\d p
\int \d q\, v\big(\varg \big)\partial_pK(p,q)\cdot u(q)\cdot\chi^\perp_{L_{1}}(q)\,
\chi^\perp_{L_{2}} ( p-\rho q)
\\&
:=\d J_{31}+\d J_{32}+\d J_{33}\;.
\end{equa}
The critical term is $\d J_{33}$, but we first deal with the two others
which are treated similar to earlier cases.

For the first term we have by Lemma~\ref{eight} which tells us that
$K$ is exponentially bounded on $\DD'$:
$$
\int \big|\d J_{31}\big|\le \const \|u\|_{{\rm L}^{\infty}}\;
\int_{|s|>(L_{2}-\HALF )/(1+\rho)}|\d v(s)|\;.
$$
where $\DD'$ is the domain
$$
  \DD'=\{ (p,q)~:~ |q|>L_1  \text{ and } |p-\rho q| > L_2\}~.
$$
For the second term, we have, again by Lemma~\ref{eight} below,
$$
\int \big|\d J_{32}\big|\le \const \|u\|_{{\rm
    L}^{\infty}}\,\|v\|_{{\rm
    L}^{\infty}}
$$
The last term is more delicate, and uses the property
$Z\cdot\lim_{p\to\pm\infty}v(p)<1$ of the definition of the
cone $vcone$, Eq.\eref{e:thecone}.
integrate by parts as before using \eref{truc} and get
\begin{equa}
\d J_{33}
=&\rho  Q\,\d p
\int 
  \d v\big(\varg \big){{\textstyle{\frac{-\rho }{1+\rho }}}}K(p,q) u(q)\cdot\chi^\perp_{L_{1}}(q)\,
\chi^\perp_{L_{2}}(p-\rho q) \\  
&\rho  Q\,\d p
\int \d q\,
  v\big(\varg \big)K(p,q) u(q)\cdot\partial_q\big(\chi_{L_{1}}(q)\chi^\perp_{L_{2}}(p-\rho q)\big)\,
\\
&\rho  Q \d p
\int  v\big(\varg \big)K(p,q)\d u(q)\cdot\chi^\perp_{L_{1}}(q)\,\chi^\perp_{L_{2}}(p-\rho q)\,
\\
:=&\,\d J_{331}+\d J_{332}+\d J_{333}\;.
 \end{equa}

The term $\d J_{331}$ is bounded like $\d J_{31}$.

In a similar way $\d J_{332}$ and $\d J_{32}$ are bounded by the same
methods.

The term $\d J_{333}$ makes use of the limit condition in $\vcone$.
Consider the integral of $|\d J_{333}|$.
This leads to a bound and setting
$L_2'=(L_2-\HALF)/(1+\rho )$:
\begin{equa}\label{positive}
\int |\d J_{333}(p)| \le\,& \rho\,Q \sup_{|s|>L_2'} 
|v(s)|\cdot \int |\d u(q)|\\&~~\cdot  \int \d
p\, K(p,q)\cdot\chi^\perp_{L_{1}}(q)\,\chi^\perp_{L_{2}}(p-\rho
q)\,\\
\le\,& \rho\,Q \left(\sup_{|q|>L_1-\HALF} \int \d
p\,K(p,q)\right)\cdot\sup_{|s|>L_2'}
|v(s)| \int |\d u| \\
=\,&\frac{\sqrt{\pi}}{\int \d q e^{-\aq} v(q)} \cdot\sup_{|s|>L_2'}
|v(s)| \int |\d u|\\
=\,&Z \cdot\sup_{|s|>L_2'}
|v(s)| \int |\d u|~,
\end{equa}
where $Z$ was defined in
Eq.\eref{Z}. 
Collecting all the estimates, we get
$$
\int \big|\d K_{v}(u)\big|\le C\int e^{-\mu q^2}|u(q)|\d q
+\zeta\big(L_{2}\big) \int |\d u|
$$
where
\begin{equa}
\zeta\big(L_{2}\big)  
&=\Oun e^{-C_2
 L_{2}^{2}}\|v\|_{\vbanach_{2}}
+\Oun\int_{|s|>L_{2}'}|\d v(s)|+Z\cdot\sup_{|s|>L_{2}'}|v(s)|~.
\end{equa}

Since $v$ belongs to $\vcone$, it follows that
$$
\lim_{L_{2}\to\infty} \zeta\big(L_{2}\big)<1\;,
$$
and the Lemma follows by taking $L_{2}$ large enough. 
\end{proof}

\begin{proposition}
For any $v\in\vcone$,
 the equation  $K_{v}(u)=u$ has a solution in $\banach_{2}$. This solution
can be chosen positive, it is then unique if we impose
$\|u\|_{\banach_{1}}=1$. We call it $u_{v}$. The map $v\mapsto u_{v}$
is differentiable. 
\end{proposition}

\begin{proof}
We apply the theorem of Ionescu-Tulcea and Marinescu \cite{TulceaMarinescu1950} to
prove the existence of $u$. Since for $v>0$, the operator $K_{v}$ is
positivity improving, it follows by a well known argument, see {\it{e.g.}},
\cite{Schaefer1999}
that the peripheral spectrum consists only of  the simple eigenvalue
one and the eigenvector can be chosen positive. If normalized, it is
then unique. Since the operator $K_{v}$ depends linearly and
continuously on $v$ (in $\vbanach_{2}$), the last result follows by
analytic perturbation theory (see \cite{Kato1984}).
\end{proof}

We next consider the equation \eref{champ} for  $v$:
\begin{equation}\label{lav}
\partial_{x}v(p)=\sign(p)\left(\frac{1}{\rho}\int
e^{(1-\rho^{2})(p-q/(1+\rho))^{2}/\rho^{2}}v\big((1-\rho)q-p)/\rho\big)
\,u_v(q)\,\d q\right.
$$
$$
\left. -v(p)\int e^{-\mu q^{2}}u_v(q)\,\d q\right)~.
\end{equation} 

\begin{proposition}
The r.h.s.~of the equation for $v$ is a $C^{1}$ vector field on
$\vbanach_{2}$. 
\end{proposition}
\begin{proof}
This follows easily from the fact that the map $v\mapsto u_{v}$ is
$C^{1}$. 
\end{proof}

\begin{theorem}
Let $v_{0}\in \vcone$, and assume that $v_{0}$ is bounded below
away from zero and has nonzero limits at $\pm\infty$. Then 
there is a number $s=s(v)>0$ such that the solution of equation
(\ref{lav}) with initial condition $v_{0}$ 
exists in $\vbanach_{2}$ and is nonnegative (moreover, it
belongs to $\vcone$).
\end{theorem}
\begin{proof}
Follows at once from the previous proposition and the fact that
$v_{0}$ is in the interior of $\vcone$.
\end{proof} 

The proof of Theorem~\ref{thm:main} is now completed by observing that
the map $\Phi: v_0\mapsto \Phi(v_0)$ is indeed a local diffeomorphism, since it is
given as the solution of an evolution equation.

\section{Remarks and Discussion}

\subsection{The behavior of the solution at $p=\infty$}
Consider the limit $p\to\infty$ in the expression for $K_{v}$. We need 
$\rho p-q=\Oun$ otherwise the Gaussian gives a negligible
contribution. In other words, $q\sim \rho p$, and we are going to
assume from now on that $\rho>0$ (the other case can be treated
analogously).  This implies 
$p-\rho q\sim (1-\rho^{2})p$ which also tends to infinity (the same
infinity). Therefore,
$$
K_{v}u(\pm\infty)=\frac{\sqrt\pi\;v(\pm\infty)\;u(\pm\infty)}{\int
  e^{-p^{2}}v(p)\d p }\;.
$$ 
In particular, if $K_{v}u=u$ and since we assumed
$$
\frac{\sqrt\pi\;v(\pm\infty)}{\int
  e^{-p^{2}}v(p)\d p }\neq 1
$$
we get $u(\pm\infty)=0$. 

For the $v$ equation, we have for large $p$, $q\sim p(1+\rho)$
and $(1-\rho)q-p\sim -\rho^{2}p$.
Therefore (inverting limit and derivative)
we get
$$
\partial_{x}v(\pm\infty)=\sign(\pm\infty)\left[
\sqrt\pi\; \sqrt{\frac{1+\rho}{1-\rho}}u(\pm\infty)\;v(\mp\infty)
-v(\pm\infty)\int e^{-\mu q^{2}}u(q)\d q\right]\;.
$$
Note that the first term vanishes since $u(\pm\infty)=0$. 
Since the integral $C(x)=\int e^{-\mu q^2 }u(q,x)$ is positive, we conclude
that formally,
$$
\partial_{x}v(\pm\infty)=\mp v(\pm\infty) C(x)~.
$$

\subsection{Essential spectrum}

\noindent\textbf{Conjecture.} \textsl{The essential spectrum of $K_{v}$
  is the interval $[0,\sigma(v)]$ with}
$$
\sigma(v)=\max\frac{\sqrt\pi\;v(\pm\infty)}{\int
  e^{-p^{2}}v(p)\d p }\;.
$$ 
If $\sigma(v)<1$ we are looking for an eigenvalue $1$ outside the
essential spectrum, which is the case we have treated.
If $\sigma(v)>1$ we would be looking for an eigenvalue
$1$ inside the essential spectrum which would be a much more difficult task,
since it may well not exist.

\noindent \textbf{Idea of proof}: Similar to the above estimates, the operator
$K_{v}$ should be written as something small plus something compact
plus something whose essential spectrum can be computed. This last
part is likely to be the limit operator at infinity.

\subsection{Dependence on $N$}\label{dep}
It should be noted that the equation for $\partial_x F$ has, in fact a
scaling of the form
$$
N^{-1}\partial_x F =\OO(1) +\OO(N^{-1})~.
$$
This means that in the main theorem (Theorem \ref{thm:main}), the
limit $x_{v _0}$ of $x$ for which we have a result is quite probably bounded by a
quantity of the form $1/(N \cdot \Delta(v _0))$, where $\Delta(v _0)$ measures the
deviation of the initial condition $v _0$ from a Gaussian. Thus,
either $x_{v _0}$ is very small when $N$ is large, or one has to
take $v _0$ very close to a Gaussian.

Another way to look at this scaling is to introduce a scattering
probability $\gamma=b/N$ where $b>0$ is a constant independent of $N$. In
other words, a particle entering the array of cells from the left has
for large $N$ a probability $e^{-b}$ to traverse all the $N$ cells
(and leave on the right) without having experienced any scattering. This is
analogous to  a rarefied gas. It is easy to verify that equation
\eref{1bg} is modified by a factor $b/N$ multiplying the right hand
side, and hence equation \eref{flfr} is unchanged. The stationary
equations \eref{eqstat} become

$$
\atpdm\flm (t,\tilde p)=\theta(-\tilde p)\left(1-\frac{b}{N}\right)
\atpdm f (t,\tilde p)
$$
$$
+\frac{b}{N}
\frac{\theta(-\tilde p)}{1-\rho }\int_{\real} \d p\,
g(t-mL/|\tilde p|,\gargf)\apdm\,f 
(t-{\textstyle{\frac{m}{|\tilde p|}}}L-\amdp L,p)\;,
$$
and
$$
\atpdm\frp (t,\tilde p)=
\theta(+\tilde p)\left(1-\frac{b}{N}\right) \atpdm f (t,\tilde p)
$$
$$
+\frac{b}{N}
\frac{\theta(+\tilde p)}{1-\rho }\int_{\real}\d p\,
g(t-mL/|\tilde p|,\gargf)\apdm\,f
(t-{\textstyle{\frac{m}{|\tilde p|}}}L-\amdp L,p)\;.
$$
Equation \eref{eqstat1} follows as explained in Section \ref{s:continuous}  after a
rescaling of space by a factor $b$.

\subsection{Discussion}
The model presented in this paper has the nice property that one can
control the existence of a solution out of equilibrium. In particular,
this means that there is no heating up of the scatterers in the
``chain'',  when the system is out of equilibrium.

The reader should note, however, that the initial condition at the
boundary, does not allow for different temperatures in the strict
sense, only for different distributions at the ends. For example, a
function of the form
\begin{equ}
F(p,0)=
\begin{cases}
  \exp(-\alpha p^2), & \text{~if~} p>0,\\
  \exp(-\alpha' p^2), & \text{~if~} p<0,\\
\end{cases}
\end{equ}
with $\alpha \ne \alpha '$ is not covered by
Theorem~\ref{thm:main}. The reason for this failure is that we could
not find an adequate analog of Lemma~\ref{eight} for initial
conditions of this type, and therefore the bounds on the kernel
$K(p,q)$ are not good enough.


\bibliographystyle{JPE}
\bibliography{refs}

\def\Rom#1{\uppercase\expandafter{\romannumeral #1}}\def\u#1{{\accent"15
  #1}}\def\cprime{$'$} \def\cprime{$'$}
\begin{thebibliography}{1}

\bibitem{DSII}
N.~Dunford and J.~T. Schwartz.
\newblock {\em Linear operators. {P}art {II}: {S}pectral theory. {S}elf adjoint
  operators in {H}ilbert space\/}.
\newblock With the assistance of William G. Bade and Robert G. Bartle
  (Interscience Publishers John Wiley \& Sons\ New York-London, 1963).

\bibitem{EMZ2006}
J.-P. Eckmann, C.~Mej{\'{\i}}a-Monasterio, and E.~Zabey.
\newblock Memory effects in nonequilibrium transport for deterministic
  {H}amiltonian systems.
\newblock {\em J. Stat. Phys.\/} {\bf 123} (2006), 1339--1360.

\bibitem{EY2}
J.-P. Eckmann and L.-S. Young.
\newblock Nonequilibrium energy profiles for a class of 1-{D} models.
\newblock {\em Comm. Math. Phys.\/} {\bf 262} (2006), 237--267.

\bibitem{TulceaMarinescu1950}
C.~T. Ionescu~Tulcea and G.~Marinescu.
\newblock Th\'eorie ergodique pour des classes d'op\'erations non
  compl\`etement continues.
\newblock {\em Ann. of Math. (2)\/} {\bf 52} (1950), 140--147.

\bibitem{Kato1984}
T.~Kato.
\newblock {\em {P}erturbation {T}heory for {L}inear {O}perators\/}
  (Springer-Verlag, Berlin, 1984).
\newblock Second corrected printing of the second edition.

\bibitem{Schaefer1999}
H.~H. Schaefer and M.~P. Wolff.
\newblock {\em Topological vector spaces\/}, volume~3 of {\em Graduate Texts in
  Mathematics\/} (New York: Springer-Verlag, 1999), second edition.

\end{thebibliography}

\section*{Acknowledgment} We thank Ph. Jacquet for a careful reading
of the manuscript. This work was partially supported by the
Fonds National Suisse.

\makeappendix{Appendix: Bounds on $K(p,q)$}\label{s:kpq}

We study here the kernel $K$ of \eref{simplekpq}, which equals
\begin{equ}
  K(p,q)= e^{E(p,q)}~,
\end{equ}
with 
\begin{equ}\label{Kpqa}
E(p,q)\,=\,\mu \ap-\mu\aq 
-(\varg)^2=-(\rho
  p-q)^2/(1+\rho )^2~.
\end{equ}
\begin{lemma}\label{l:supkpq}
Assume $|q|<L$. There are constants $C=C(L,\rho )$ and
$D=D(L,\rho )>0$ such that for all $p$,
\begin{equ}\label{e:supkpq}
  K(p,q) < C e^{-Dp^2}~,
\end{equ}
and
\begin{equ}\label{e:supdkpq}
|\partial_p  K(p,q)| < C e^{-Dp^2},
\end{equ}
\end{lemma}
\begin{proof}
Obvious.
\end{proof}
\begin{lemma}\label{l:supkpq2}
Assume $|p-\rho q|<L$. There are constants $C=C(L,\rho )$ and
$D=D(L,\rho )>0$ such that for all $q$,
\begin{equ}\label{e:supkpq2}
  K(p,q) < C e^{-Dq^2}~,
\end{equ}
and
\begin{equ}\label{e:supdkpq2}
|\partial_p  K(p,q)| < C e^{-Dq^2},
\end{equ}
\end{lemma}
\begin{proof}
  The proof is as in  Lemma~\ref{l:supkpq}, with the
  difference that now $|p-\rho q|<L$. 
\end{proof}

\begin{lemma}\label{l:kpq} Consider the domain $\DD$ defined by
  \begin{equ}\label{dd}
    \DD= \{ (p,q)\in \real^2~:~ |p-\rho q|< L_2+\HALF\text{ and } |q|
    > L_1-\HALF\}~,
  \end{equ}
with 
\begin{equ}\label{L1L2}
L_{1}=\frac{3\rho}{1-\rho^{2}}L_{2}\;.
\end{equ}

For fixed $\rho \in (0,1)$ and sufficiently large $L_2$ there are
positive constants $C_1$ and $C_2$ such that for 
$(p,q)\in \DD$ one has the bound
\begin{equ}
  K(p,q)< \exp\left(-C_1 (\rho
  p-q)^2-C_2 L_2^2\right)~.
\end{equ}
\end{lemma}
\begin{proof}
From the definition of $\DD$ and $(1-\rho
^2)q =(\rho p-q)- \rho (p-\rho q)$, we find (for sufficiently large $L_2$):
\begin{equ}\label{L111}
|\rho p-q|\ge \big(1-\rho^{2}\big)|q|-\rho |p-\rho q|\ge 
\big(1-\rho^{2}\big)\big(L_{1}-\HALF \big)-\rho\big(L_{2}+\HALF \big)>
\rho L_2\;.
\end{equ}
Using  the form
\begin{equ}\label{L111split}
  (\rho p-q)^2 > \FOUR(\rho p-q)^2 +\FOUR\rho^2 L_2^2~,
\end{equ}
the assertion follows immediately.
\end{proof}

We next study the region
\begin{equ}\label{badregion}
  \DD'=\{ (p,q)~:~ |q|>L_1  \text{ and } |p-\rho q| > L_2\}~.
\end{equ}
In this region, we have the obvious bound
\begin{lemma}\label{eight}
For $(p,q)\in\DD'$,
one has the bound 
$$
E(p,q) = -(\ttt{\frac{\rho p -q}{1+\rho }})^2~.
$$
 
\end{lemma}

\end{document}